\documentclass[twocolumn,aps,prl,floatfix,showpacs,amsmath,amssymb]{revtex4}

\usepackage{epsfig}
\usepackage{graphicx}
\usepackage{dcolumn}
\usepackage{bm}

\begin{document}
\title{Finite-time scaling of dynamic quantum criticality}
\author{Shuai Yin} \email{zsuyinshuai@163.com}
\author{Xizhou Qin}
\author{Chaohong Lee} \email{lichaoh2@mail.sysu.edu.cn}
\author{Fan Zhong}  \email{stszf@mail.sysu.edu.cn}

\affiliation{State Key Laboratory of Optoelectronic Materials and
Technologies, School of Physics and Engineering, Sun Yat-sen
University, Guangzhou 510275, People's Republic of China}

\begin{abstract}
We develop a theory of finite-time scaling for dynamic quantum
criticality by considering the competition among an external time
scale, an intrinsic reaction time scale and an imaginary time scale
arising respectively from an external driving field, the
fluctuations of the competing orders and thermal fluctuations.
Through a successful application in determining the critical
properties at zero temperature and the solution of real-time
Lindblad master equation near a quantum critical point at nonzero
temperatures, we show that finite-time scaling offers not only an
amenable and systematic approach to detect the dynamic critical
properties, but also a unified framework to understand and explore
nonequilibrium dynamics of quantum criticality, which shows specificities for open systems.
\end{abstract}

 \maketitle

\date{\today}

Detecting quantum phase transitions (QPTs) and understanding their
real-time dynamics are of great importance
\cite{sachdev,Coleman,sachdevpt,Dziarmaga,polrmp}. Recent
experimental breakthrough in ultracold atoms \cite{Greiner} promises
new tools to study the quantum critical dynamics \cite{Zhang}. In
the nonequilibrium critical dynamics of QPTs at zero temperature in
which a controlling parameter is changed with time through a
critical point \cite{Dziarmaga,polrmp}, the Kibble-Zurek mechanism
(KZM), which was first introduced in cosmology by Kibble
\cite{kibble1} and then in condensed matter physics by Zurek
\cite{zurek1}, has been found to describe the dynamics of QPTs well
\cite{Dziarmaga,polrmp,Zurek}. In this adiabatic--impulse--adiabatic
approximation of the KZM, the system considered is assumed to cease
evolving in the impulse regime within which adiabaticity breaks down
due to critical slowing down \cite{binder}. Yet, dynamical scaling
has been reported just within this regime \cite{deng} and confirmed
in both integrable and nonintegrable systems
\cite{grandi,kolodrubetz}. In classical critical dynamics, an
explanation based on coarsening has been developed
\cite{biroli}, however in quantum phase transitions, a systematic
understanding of the full scaling behavior is still lacking.

On the other hand, natural systems and their measurements exist
inevitably in nonzero temperatures, though probably only initial
thermal states need considering in ultracold atoms
\cite{Polkovnikov}. Thermal effects on a quantum critical state can
give rise to a variety of exotic behavior in the famous quantum
critical regime (QCR)~\cite{Chakravarty} as exhibits in a wide range
of strongly correlated systems
\cite{sachdev,Coleman,sachdevpt,Broun}. Yet, as both phases exhibit
complex long-range quantum entanglement near the quantum critical
point and are violently excited thermally, it is a great challenge
to describe quantum critical dynamics at finite temperatures, let
alone nonequilibrium real-time effects \cite{patane,chandran}. Indeed, none
of the analytic, semiclassical, or numerical methods of
condensed-matter physics yields accurate results for dynamics in the
QCR except for some special systems in 1D \cite{sachdevpt}. Even in isolated situations it is difficult to study the time evolution of nonequilibrium
systems with many degrees of freedom
\cite{Kinoshita,Hofferberth,rigol,Dziarmaga,polrmp}. Therefore,
systematic approaches have to be invoked.


\begin{figure}
  \centering
   \includegraphics[bb= 0 0 362 364, clip, scale=0.4]{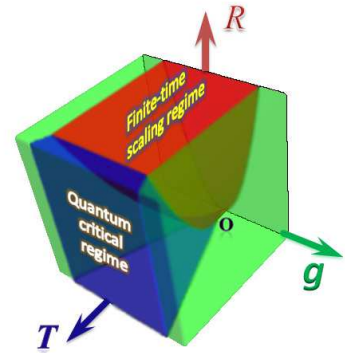}
   \caption{\label{ftsqcr}(color online) Schematic phase diagram under a sweep of $g$ near its critical value 0. Two equilibrium phases (light grey/green) dominated by the reaction time $\tau_s$ are separated by two crossover domains fanning out from the
   quantum critical point $O$. One domain is the QCR (dark/blue) controlled by the imaginary time scale $\tau_T$. The other is the new FTS (grey/red) regime governed by the driving time scale $\tau_d$.}
\end{figure}
Time plays a fundamental role in quantum criticality owing to the
interplay of static and dynamic behaviors. Specifically, by varying the distance to the critical point $g$ at a time rate $R$, a continuous QPT
at a finite temperature $T$ is characterized mainly by three time scales.
The first one is a reaction time $\tau_s$ that arises from the
fluctuations of the competing orders and blows up as $\tau_s\sim
|g|^{-\nu z}$ with the standard critical exponents $\nu$ and $z$ as
$g$ vanishes \cite{sachdev}. The
second one is an ``imaginary'' time scale $\tau_T=1/T$ (the Plank
and the Boltzmann constants have been set to 1) due to the finite
$T$, since the real time is its analytical continuation to imaginary
numbers through a Feynman path integral representation
\cite{sachdev}. The third one is an externally imposed driving time scale
$\tau_d$ that results from the driving and grows as
$\tau_d\sim R^{-z/r}$ with a rate exponent $r$ that is related to
$z$ and other static critical exponents
\cite{Gong}. It is the competition among $\tau_s$,
$\tau_T$, and $\tau_d$ that lead to a diversity of equilibrium and
dynamic universal phenomena near a quantum critical point.

Here we study systematically the competition among the three
characteristic time scales according to the theory of finite-time
scaling (FTS) \cite{Gong}. As seen in Fig.~\ref{ftsqcr}, besides the
usual equilibrium regimes and the QCR which are respectively
dominated by $\tau_s$ and $\tau_T$, our most important result is
that a new nonequilibrium FTS domain is created. In this domain,
$\tau_d$ is the shortest time among the trio and thus dominates,
just as the well-known regime of finite-size scaling in which the
characteristic size $L$ of the system is shorter than its
correlation length. At $T=0$, this indicates the FTS domain overlaps
just the impulse regime of the KZM for sweeping $g$. As a
consequence, although the system falls out off equilibrium, the
state does not cease evolving; rather, it evolves according to the
imposed time scale $\tau_d$ instead of $\tau_s$ with nonadiabatic
excitations obeying FTS. Therefore, FTS improves the understanding
of KZM on its dark impulse regime and produces naturally scaling
forms suggested in \cite{deng,grandi,kolodrubetz}. In addition, FTS
enables us to the study within the same framework other driving
dynamics than the KZ protocols~\cite{chandran}, which focus on
changing non-symmetry breaking terms like $g$. We shall show that
these provides a convenient method to determine the critical points
and exponents, which were invoked as input for scaling collapses
reported in \cite{deng,grandi,kolodrubetz}. Similarly, at $T\neq0$,
in the FTS regime, there are now nonadiabatic thermal excitations
controlled again by $\tau_d$ and thus obeying again FTS. Further,
from Fig.~\ref{ftsqcr}, the FTS regime pushes the QCR to higher
temperatures since only then $\tau_T$ dominates. Consequently, FTS
enables one to probe directly the quantum critical point and its
scaling behavior at nonzero temperatures as $T$ becomes subordinate
and just a perturbation. Thus, it shows a `dynamic cooling' effect
that enables one to probe the zero-temperature scaling at
nonzero-temperatures while keeping $T$ subsidiary. This offers us an
extra approach to detect and study quantum criticality at finite
temperatures. Owing to its conceptual simplicity and accessability,
FTS therefore provides a unified framework not only to detect the
dynamic critical properties, but also to understand and explore the
nonequilibrium dynamics of quantum criticality both at $T=0$ and
$T\neq0$.

As another important result, we shall show that in nonequilibrium quantum critical dynamics of open systems one must include an additional variable such as the coupling to a heat bath to the intrinsic quantum dynamics~\cite{note}. This is an important difference from the classical case and must be considered when extending nonequilibrium quantum critical dynamics to finite temperatures~\cite{patane}.
We shall see that the master equation in the Lindblad form just offers such a variable and is thus an appropriate platform to study real-time nonequilibrium quantum criticality.

We start with an open many-body quantum system interacting with a
heat bath \cite{patane} to study the interplay of quantum and
thermal fluctuations. The state of such a system can be described by
a density matrix operator $\rho$ according to quantum statistical
physics. For weak system-environment couplings, after assuming
Markovian and tracing over the bath variables, one obtains the
master equation for $\rho$ in the Lindblad form
\cite{Lindblad,attal,Mai},
\begin{equation}
\partial \rho/\partial t=-i[\mathcal H,\rho]+c\mathcal L\rho
\label{Lind},
\end{equation}
where $\mathcal L\rho=-\sum_{i=1,j\neq
i}^{N_{E}}\beta_{j}(V_{i\rightarrow j}^{\dag}V_{i\rightarrow
j}\rho+\rho V_{i\rightarrow j}^{\dag}V_{i\rightarrow j}
-2V_{i\rightarrow j}\rho V_{i\rightarrow j}^{\dag})/2$, $c$ is the
dissipation rate and measures the coupling strength between the
system and the bath, $N_{E}$ is the total number of energy levels,
$\beta_{i}=\textrm{exp}(-E_{i}/T)/\textrm{Tr} \textrm{exp}(-\mathcal
H/T)$ with $E_{i}$ being the $i$th eigenvalue of $\mathcal{H}$, and
$V_{i\rightarrow j}$ is the thermal jump matrix whose element at the
$j$th row and $i$th column is one or zero in the energy
representation. $V_{i\rightarrow j}$ fulfills $\beta_{i}\rho_E
V_{i\rightarrow j}=\beta_{j}V_{i\rightarrow j }\rho_E$ with the
equilibrium density matrix operator
$\rho_E\equiv\textrm{exp}(-\mathcal H/T)/\textrm{Tr}
\textrm{exp}(-\mathcal H/T)$ whose eigenvalues are $\beta_{i}$. This
can be regarded as a detailed balance condition in equilibrium. The
Lindblad equation~(\ref{Lind}) is a real-time dynamical equations
which integrates both the quantum and the thermal contributions. It
has been widely used in quantum optics \cite{orszag} and relaxation
processes in open quantum systems \cite{attal,znidaric}. Although
for large couplings, Eq.(\ref{Lind}) may be
inapplicable~\cite{wgwang}, this equation gives a reasonable
description in the weak coupling limit, for instance, for the
time-independent $\mathcal H$, the steady solution of Eq.
(\ref{Lind}) is $\rho_E$ independent of $c$
\cite{Lindblad,attal,Mai}, this is consistent with the foundation of
statistical mechanics \cite{Schrodinger}. In the following, we focus
on the weak coupling limit and consider the scaling properties of
the Lindblad equation.

The theory of FTS \cite{Gong} takes explicitly into account the rate
$R$, which plays a role similar to $L^{-1}$ since it imposes on a
system an additional time scale that manipulates its evolution. In
classical critical dynamics, the nonequilibrium dynamic scaling
can be generalized directly from the equilibrium ones as confirmed by the renormalization-group theory \cite{Gong}. However, in the nonequilibrium quantum
criticality, as pointed out, a coupling strength must be considered as an independent
scaling variable. In the weak coupling limit, this strength can be
reduced to the dissipation rate $c$. Accordingly, for a length rescaling of
factor $b$, an order parameter $M$ transforms as
\begin{eqnarray}
M(t,g,h_z,T,L,c,R)= b^{-\beta /\nu}M(tb^{-z}, gb^{1/\nu},h_zb^{\beta\delta/\nu}, \quad\nonumber\\
Tb^{z}, L^{-1}b,cb^{z}, Rb^{r}),\quad\label{opp}
\end{eqnarray}
where the two critical exponents $\beta$ and $\delta$ are defined as
usual in classical critical phenomena by $M\propto g^\beta$ in the
absence of an external probe field $h_z$ conjugate to $M$ and
$M\propto h_z^\delta$ at $g=0$, respectively. In the weak coupling
limit, $c$ is small thus one can expect its scaling behavior is
controlled by the the fixed point corresponding to the critical
point at $c=0$, thus the dimension of $c$ is identical with $t^{-1}$
as can be inspected from Eq.~(\ref{Lind}) \cite{Mai}. This is
checked latter by the numerical solution of Eq. (\ref{Lind}).

With Eq.~(\ref{opp}), one can describe in a unified framework
different kinds of driven dynamics via changing $g$, $h_z$ or $T$
and readily define different regimes and their crossovers. Taking
$g=Rt$ for instance, neglecting $h_z$, suppressing one independent variable, and choosing $b$ such that
$Rb^{r}$ becomes a constant, one finds an FTS scaling form
\begin{eqnarray}
M(g,T,L,c,R)= R^{\beta /\nu r}f_1(gR^{-1/\nu r},TR^{-z/r},\quad\nonumber\\
L^{-1}R^{-1/r},cR^{-z/r}),\ \label{opt}
\end{eqnarray}
where $r=z+1/\nu$ obtained from $g=Rt$ and its rescaling \cite{Gong}
and the function $f_i$ with an integer $i$ denotes a scaling
function. FTS dominates when $|g|R^{-1/\nu r}\ll1$, $TR^{-z/r} \ll
1$, $L^{-1}R^{-1/r} \ll 1$, and $cR^{-z/r} \ll 1$. The first gives
$\tau_d\sim R^{-z/r}\ll |g|^{-\nu z}\sim\tau_s$, the second
$\tau_d\ll1/T=\tau_T$ as they ought to be. Crossovers to other
regimes occur near $|\hat{g}|\sim R^{1/\nu r}$ and $\hat{T} \sim
R^{z/r}$ as depicted in Fig.~\ref{ftsqcr} and similar ones for $L$
and $c$.
The first gives $|\hat{t}|\sim R^{-\nu z/(1+\nu z)}$ because
$\hat{g}=R\hat{t}$. This is just the scaling of the KZM upon
identifying $\hat{t}$ with the freeze-out time instant
\cite{Dziarmaga,polrmp,Zurek} for a closed system $c=0$ in the
thermodynamic limit ($L\rightarrow\infty$) and at $T=0$.

Several remarks are in order here. (a)~Equation~(\ref{opt}) is different from the similar scaling form for finite temperatures in \cite{chandran}
because $c$ must be included to introduce the
thermal fluctuation in the nonequilibrium situation. (b)~To return to the equilibrium scaling form at finite-temperatures
\cite{sachdev}, the scaling function $f_i$ must satisfy a constraint
of $\partial f_i/\partial c=0$ for $R=0$. (c)~Beside recovering
the full scaling forms of finite-size for closed system in
\cite{grandi,kolodrubetz} by fixing $c=0$ and $T=0$, the nonequilibrium dissipation
scaling for spontaneous emissions in
zero-temperature open quantum systems can also be studied by fixing $T=0$ in
Eq.~(\ref{opt}). (d)~Note that $c$ should be small in the weak coupling limit and thus the regime dominated by $c$ may be inaccessible.

Instead of sweeping $g$, when $h_z=R_zt$, one obtains similarly the order parameter
\begin{eqnarray}
M_h=R_z^{\beta /\nu r_z}f_2(gR_z^{-1/\nu
r_z},h_zR_z^{-\beta \delta /\nu
r_z},\qquad\quad\nonumber\\
TR_z^{-z/r_z},L^{-1}R_z^{-1/r_z},cR^{-z/r_z})\!\!\!\label{op2}
\end{eqnarray}
with $r_z=z+\beta\delta/\nu$. Different regimes and their crossovers
can also be readily defined. Different from sweeping $g$ through the
critical point as the ordinary KZM protocols
\cite{deng,grandi,kolodrubetz,chandran}, here we fix $g$ and change
the symmetry breaking field $h_z$. This provides a method to
determine the critical point from distinct critical behaviors for
$g=0$ and $g\neq0$, a method which we shall utilize below and may
also be realizable experimentally. Note that in this protocol, the
form of $\tau_d$ remains remarkably if $R$ and $r$ are replaced with
their counterparts. However, in addition to the fixed $\tau_s$ for
the fixed $g$, there exists another reaction time diverging with
$|h_z|^{-\nu z/\beta \delta}$. These result in new competitions but
act only as corrections in the FTS regime, showing an advantage of
FTS.

Now we show that FTS can provide methods to detect quantum critical
properties such as the critical point and critical exponents. For simplicity, we consider $T=0$ and $c=0$ in the thermodynamic limit
$L\rightarrow\infty$. According to Eq.~(\ref{op2}), at $h_z=0$, $M_h$ reduces to
\begin{equation}
M_0(g,R_z)=R_z^{\beta /\nu r_z}f_3(gR_z^{-1/\nu r_z}), \label{op3}
\end{equation}
while the field at $M_h=0$, denoted by $h_{z0}$, scales as
\begin{equation}
h_{z0}(g,R_z)=R_z^{\beta \delta /\nu r_z}f_4(gR_z^{-1/\nu r_z}).
\label{ef}
\end{equation}
Differentiating $M_h$ with respect to $h_z$ in Eq.~(\ref{op2}), one
obtains the susceptibility at zero field,
\begin{equation}
\chi(g,R_z)=R_z^{\beta (1-\delta)/\nu r_z}f_5(gR_z^{-1/\nu r_z}).
\label{sus}
\end{equation}
To fix the critical point, we can define a cumulant
\begin{equation}
C(g,R_z)\equiv M_0/(h_{z0}\chi)=f_6(gR_z^{-1/\nu r_z}) \label{cum}
\end{equation}
similar to the Binder cumulant in finite-size scaling~\cite{binder}.
As $C$ is a function of only one independent variable, its curves
for different $R_z$ intersect at the critical point $g=0$ at which
$C$ becomes a constant $f_6(0)$ independent on $R_z$. This gives the
critical point with which all the critical exponents can then be
estimated. For example, $\beta /\nu r_z$ and $\beta \delta /\nu r_z$
can be estimated respectively from Eqs.~(\ref{op3}) and (\ref{ef})
by fitting $M_0$ and $h_{z0}$ for a series of $R_z$ at $g=0$.
Similarly, from Eq.~(\ref{opt}) at $c=0$, $T=0$, and $L\rightarrow\infty$, $\beta /\nu
r$ can be estimated by fitting $M$ for a series of $R$ at $g=0$.
From these three exponent ratios and the scaling law~\cite{sachdev}
$\beta(\delta +1)=(d+z)\nu$ with the space dimensionality $d$, one
can determine all the critical exponents.

As an example of the FTS method to determine critical properties, we
consider the one-dimensional (1D) transverse-field Ising model whose
Hamiltonian is~\cite{sachdev}
\begin{equation}
\mathcal {H}=-h_x\sum\limits_{n=1}^N\sigma_n
^x-\sum\limits_{n=1}^{N-1}\sigma_n^z\sigma_{n+1}^z, \label{Hamil1}
\end{equation}
and has been realized in CoNb$_2$O$_6$ experimentally \cite{Coldea},
where $\sigma_n^x$ and $\sigma_n^z$ are the Pauli matrices, $h_x$ is
the transverse field, and the Ising coupling has been set to unity
as our energy unit. The model exhibits a continuous QPT from a
ferromagnetic phase to a quantum paramagnetic phase at a critical
point $h_{xc}$ (and so $g=h_x-h_{xc}$) at $T=0$ \cite{sachdev}. The
order parameter is the magnetization
$M=\sum_{n=1}^N\langle\sigma_n^z\rangle/N$ for the $N$ spins with
the angle brackets denoting the quantum and/or thermal average. As a
method to probe the transition, we add to $\mathcal{H}$ a
symmetry-breaking term $-h_z\sum_{n=1}^N\sigma_n ^z$.

\begin{figure}
\centerline{\epsfig{file= 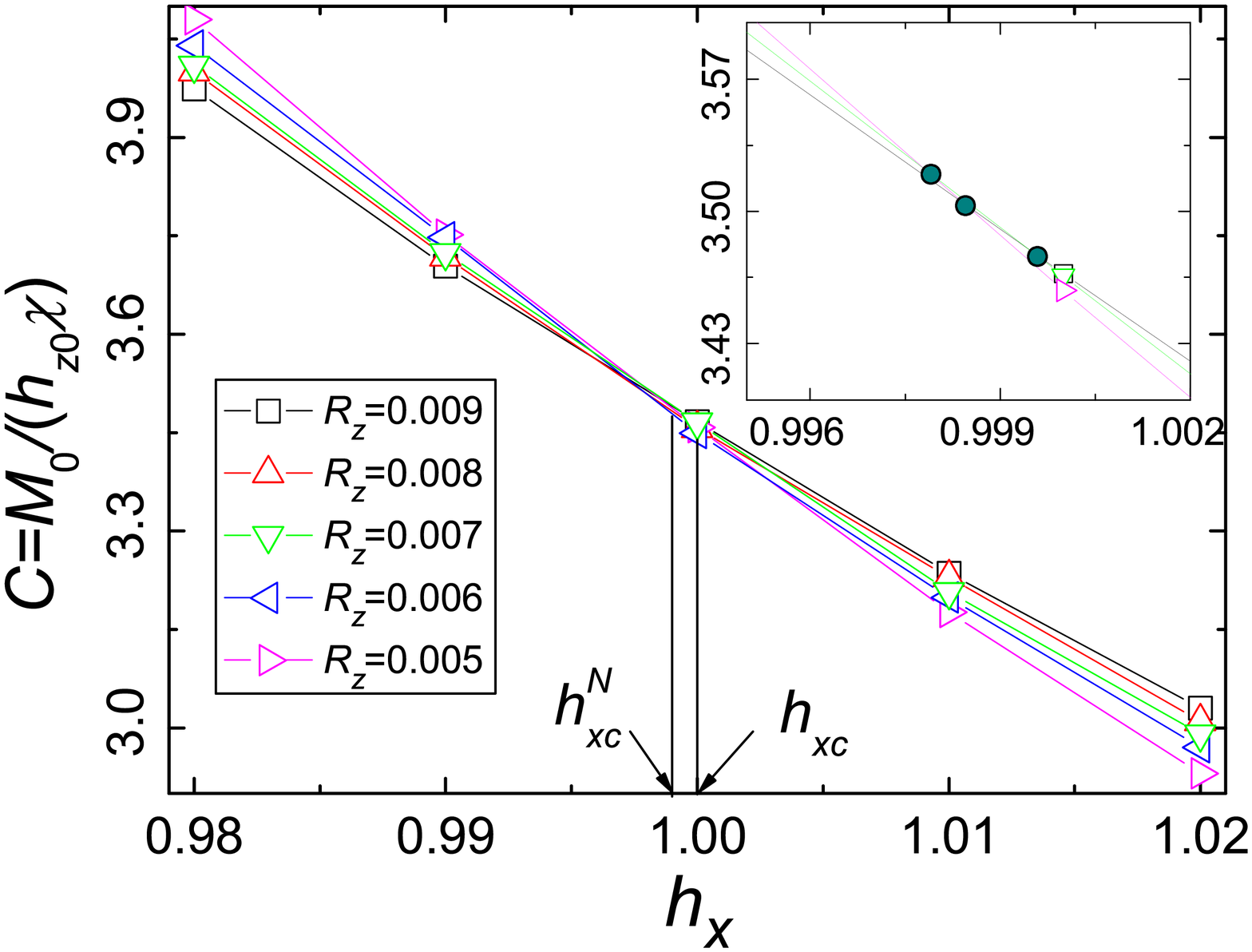,width=0.82\columnwidth}}
\caption{\label{cp} (color online) Estimation of quantum critical
point. Curves of the cumulant $C$ for different $R_z$ intersect at
the critical point $h_{xc}$ or $g=0$. Owing to possible errors from
the truncation of the singular values in the Schmidt
decomposition~\cite{Vidal}, however, the intersections are slightly
scattered as shown in the inset. Nevertheless, the average of all
the intersections is $h_{xc}^N=0.999(2)$, a good estimate of the
exact value $h_{xc} =1$. We choose a lattice size of $L=2000$, which
has been checked to produce a negligible size effect.}
\end{figure}

\begin{figure}
\centerline{\epsfig{file=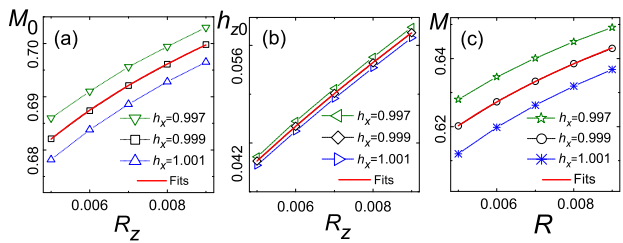,width=1.06\columnwidth}} 
\caption{\label{op} (color online) Estimation of critical exponents.
Our solutions with $h_{xc}^N =0.999$, $T=0$, and $L=2000$ yield
$\beta /\nu r_z = 0.0436$, $\beta \delta /\nu r_z=0.651$, and $\beta
/\nu r= 0.0622$ from power-law fits according to the scaling
forms~(\ref{op3}), (\ref{ef}) and (\ref{opt}), respectively. We then
obtain all the critical exponents listed in Table~\ref{table1} with
their exact results for comparison. As the statistical errors of the
fits are tiny, we fit data at $h_{xc} =0.997$ and $1.001$ and the
largest difference in each exponent is used as an estimate of the
error given also in Table~\ref{table1}.}
\end{figure}

We illustrate our approach at $T=0$ and $c=0$ at which
Eq.~(\ref{Lind}) is same to Sch\"{o}dinger's equation and some exact
results are available for comparison. We solve the model using the
time-evolving block-decimation algorithm \cite{Vidal}, which is
capable of treating large system sizes. We determine the critical
point in Fig.~\ref{cp} and apply it purposely to determine the
critical exponents in Fig.~\ref{op}. The good agreement of the
results collected in Table~\ref{table1} shows the power of FTS.
\begin{table}
\caption{\label{table1}Critical point and exponents for the 1D
transverse-field Ising model}
\begin{ruledtabular}
\begin{tabular}{ccccccccc}
$ $ & $h_{xc}$ & $\beta$ & $\delta$ & $\nu$ & $z$\\
\hline
Numerical & 0.999(2) & 0.125(11) & 14.9(6) & 0.98(4) & 1.01(3)\\
Exact~\protect\cite{sachdev} & 1 & 0.125 & 15 & 1 & 1
\end{tabular}
\end{ruledtabular}
\end{table}

\begin{figure}[htb]
\centerline{\epsfig{file=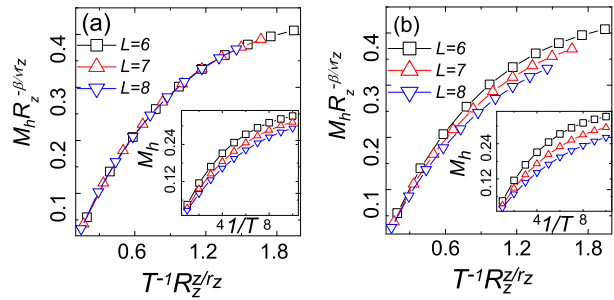, width=1.06 \columnwidth}}
\caption{\label{ropT} (color online) Nonequilibrium scaling at
nonzero temperatures. (a) Data of $M_h$ versus $T$ plotted in the
inset for the three different sets of $R_z$, $c$, and $L$ so choosing
as to fix the value of $L^{-1}R_z^{-1/r_z}$ and $cR_z^{-z/r_z}$
collapse as expected onto a single curve for the fixed
$LR_z^{1/r_z}=1.166$ and $cR_z^{-z/r_z}=3.603$ according to the
FTS~(\ref{op2}) at $g=0$ ($h_x=0.999$), $h_z=0$. (b) If, instead of $cR_z^{-z/r_z}$, we fix all $c=0.7$, the value for $L=6$, and keep others, the rescaled curves then do \emph{not} collapse.}
\end{figure}
Having successfully demonstrated FTS at $T=0$, we now turn to
$T\neq0$ at which most experiments operate. To examine the general
nonequilibrium FTS~(\ref{op2}) for $T\neq0$, we solve numerically
Eq.~(\ref{Lind}) for the Hamiltonian~(\ref{Hamil1}) along with the
field $h_z$ by a finite difference method to second order with periodic boundary conditions. We find that
$M_h$ can now saturate correctly with the thermal fluctuations.
Moreover, Fig.~\ref{ropT} shows clearly the
validity of the FTS form~(\ref{op2}). Further, upon
comparing (a) with (b) in Fig.~\ref{ropT}, it is obvious that $c$ must enter into the scaling forms with a scaling dimension $z$. Note that
although here we only solve directly Eq.~(\ref{Lind}) for small
lattices, the results show that it is suitable for describing the
nonequilibrium behavior at finite temperatures near the quantum
critical point. Moreover, the rapidly developing numerical
renormalization-group methods \cite{cirac}, for example, seem quite
promising to solve the equation for larger lattice sizes
\cite{cirac2}.

In conclusion, FTS not only provides a unified understanding of the
driving dynamics in general and lights up the dark impulse regime of
KZM at zero temperature in particular, but also sheds light on the
QCR at nonzero temperatures by establishing its own regime. It
offers a powerful unified approach amenable to both numerics and
experiments to study equilibrium and nonequilibrium dynamics of
quantum criticality. We have shown that in the latter
in open systems one must include the dissipation rate
as an independent scaling variable and the Lindblad
equation can be a valuable framework for such studies. Although we have
studied a simple model for illustration, our approach should be
applicable to more complex systems as well. In addition, as our
results indicate that the classical theory of FTS with proper
modifications can well describe quantum criticality, new physics may
be in action \cite{Coleman} if it is violated.

We thank Junhong An, Peter Drummond, and Xiwen Guan for their
valuable comments and discussions. Y.S. and F.Z. were supported by
NNSFC (10625420) and FRFCUC. C.L. was supported by NNSFC (11075223),
NBRPC (2012CB821300  (2012CB821305)), NCETPC (NCET-10-0850). We
acknowledge use of some source codes for TEBD from
http://physics.mines.edu/downloads/software/tebd/.


\begin{thebibliography}{99}

\bibitem{sachdev}S. Sachdev,  {\it Quantum Phase Transitions}, (Cambridge University Press, 1999).
\bibitem{Coleman}P. Coleman and A. J. Schofield, Nature {\bf 433}, 226 (2005).
\bibitem{sachdevpt}S. Sachdev and B. Keimer, Phys. Today {\bf 64}(2), 29 (2011).
\bibitem{Dziarmaga}J. Dziarmaga, Adv. Phys {\bf 59}, 1063 (2010).
\bibitem{polrmp}A. Polkovnikov, K. Sengupta, A. Silva, and M. Vengalattore, Rev. Mod. Phys {\bf 83}, 863 (2011).
\bibitem{Greiner}M. Greiner, {\it et al.}, Nature {\bf 415}, 39 (2002).
\bibitem{Zhang}X. Zhang, C-L. Hung, S-K. Tung, and C. Chin, Science {\bf 335}, 1070 (2012).
\bibitem{kibble1} T. Kibble, J Phys. A: Math. Gen. {\bf 9}, 1387
(1976); Phys. Today  {\bf 60}(9), 47 (2007).
\bibitem{zurek1} W. H. Zurek, Nature {\bf 317}, 505 (1985).
\bibitem{Zurek}W. H. Zurek, U. Dorner, and P. Zoller, Phys. Rev. Lett. {\bf 95}, 105701 (2005); J. Dziarmaga, Phys. Rev. Lett.
{\bf 95}, 245701 (2005); A. Polkovnikov, Phys. Rev. B {\bf 72},
161201(R) (2005); B. Damski and W. H. Zurek, Phys. Rev. Lett. {\bf
99}, 130402 (2007); F. M. Cucchietti, B. Damski, J. Dziarmaga, and
W. H. Zurek, Phys. Rev. A {\bf 75}, 023603 (2007); L. Cincio, J.
Dziarmaga, M. M. Rams, and W. H. Zurek, Phys. Rev. A {\bf 75},
052321 (2007); V. Mukherjee, U. Divakaran, A. Dutta, and D. Sen,
Phys. Rev. B {\bf 76}, 174303 (2007); D. Sen, K. Sengupta, and S.
Mondal, Phys. Rev. Lett. {\bf 101}, 016806 (2008); S. Mondal, K.
Sengupta, and D. Sen. Phys. Rev. B {\bf 79}, 045128 (2009); C. Lee,
Phys. Rev. Lett. {\bf 102}, 070401 (2009); C. De Grandi, V. Gritsev,
and A. Polkovnikov, Phys. Rev. B {\bf 81}, 012303 (2010); C. De
Grandi, V. Gritsev, and A. Polkovnikov, Phys. Rev. B {\bf 81},
224301 (2010).
\bibitem{binder}D. P. Landau and K. Binder, {\it A Guide to Monte Carlo Simulations in Statistical Physics}, 2nd edition (Cambridge University Press, Cambridge, 2009).
\bibitem{deng}S. Deng, G. Ortiz, and L. Viola, Europhys. Lett. {\bf 84}, 67008 (2008).
\bibitem{grandi}C. De Grandi, A. Polkovnikov, and A. W. Sandvik, Phys. Rev. B {\bf 84}, 224303 (2011).
\bibitem{kolodrubetz}M. Kolodrubetz, D. Pekker, B. K. Clark, and K. Sengupta, Phys. Rev. B {\bf 85}, 100505(R) (2012); M. Kolodrubetz, B. K. Clark, and D. A. Huse, Phys. Rev. Lett. {\bf
109}, 015701 (2012).
\bibitem{biroli}G. Biroli, L. F. Cugliandolo, and A. Sicilia, Phys. Rev. E {\bf 81}, 050101(R)
(2010); A. Jelic and L. F. Cugliandolo, J. Stat. Mech. P02032
(2011).
\bibitem{Polkovnikov}A. Polkovnikov and V. Gritsev, Nat. Phys. {\bf 4}, 477 (2008); S. Sotiriadis, P. Calabrese, and J. Cardy, Europhy. Lett. {\bf 87},
20002 (2009). V. Gritsev and A.  Polkovnikov, in
\textit{Understanding Quantum Phase Transitions}. ed. L. D. Carr,
(Taylor \& Francis, Boca Raton, 2010); S. Deng, G. Ortiz, and L.
Viola, Phys. Rev. B {\bf 83}, 094304 (2011).
\bibitem{Chakravarty}S. Chakravarty, B. I. Halperin, and D. R. Nelson, Phys. Rev. B {\bf 39}, 2344 (1989).
\bibitem{Broun}D. M. Broun, Nat. Phys. {\bf 4}, 170 (2008).
\bibitem{patane}D. Patan\`{e}, A. Silva, L. Amico, R. Fazio, and G. E. Santoro, Phys. Rev. Lett. {\bf 101}, 175701 (2008).
\bibitem{chandran}A. Chandran, A. Erez, S. S. Gubser, and S. L. Sondhi, Phys. Rev. B {\bf 86}, 064304 (2012).
\bibitem{note} We thank an anonymous referee for pointing out this to us.
\bibitem{Kinoshita}T. Kinoshita, T. Wenger, and D. S. Weiss. Nature {\bf 440}, 900 (2006).
\bibitem{Hofferberth}S. Hofferberth, {\it et al.}, Nature {\bf 449}, 324 (2007).
\bibitem{rigol}M. Rigol, V. Dunjko, and M. Olshanii, Nature {\bf 452}, 854 (2008).
\bibitem{Gong}S. Gong, F. Zhong, X. Huang, and S. Fan, New J. Phys. {\bf 12}, 043036 (2010); F. Zhong, in {\it Applications of Monte Carlo Method in Science and Engineering}. ed. S. Mordechai, p469 (Intech, 2011). Available at http://www.intechopen.com/books/applications-of-monte-carlo-method-in-science-and-engineering/finite-time-scaling-and-its-applications-to-continuous-phase-transitions.
\bibitem{Lindblad}G. Lindblad, Commun. Math. Phys. {\bf 48}, 119 (1976).
\bibitem{attal}S. Attal and A. Joye, J. Func. Analysis. {\bf 247}, 253 (2007).
\bibitem{Mai}P. Mai, Derivation of the Lindblad equation from a microscopic mechanism in which the open Ising chain coupling weakly with an infinite thermal bath. (unpublished).
\bibitem{orszag}M. Orszag, {\it Quantum optics}, 2nd Edition. (Springer, 2008).
\bibitem{znidaric}M. \v{Z}nidari\v{c}, T. Prosen, G. Benenti, G. Casati, and D. Rossini, Phys. Rev. E {\bf 81}, 051135 (2010).
\bibitem{wgwang}W. G. Wang, Phys. Rev. E {\bf 86}, 011115 (2012).
\bibitem{Schrodinger}E. Schr\"{o}dinger, {\it Statistical Thermodynamics} (Cambridge University Press, Cambridge, England,
1952); S. Goldstein, J. L. Lebowitz, R. Tumulka, and N. Zangh\`{\i},
Phys. Rev. Lett. {\bf 96} , 050403 (2006); J. Cho and M. S. Kim,
Phys. Rev. Lett. {\bf 104}, 170402 (2010); S. Goldstein, {\it et
al.}, Phys. Rev. E {\bf 81} , 011109 (2010); S. Popescu, A. J.
Short, and A. Winter, Nature Phys. {\bf 2}, 754 (2006).
\bibitem{Coldea}R. Coldea, {\it et al.} Science {\bf 327}, 177 (2010).
\bibitem{Vidal}G. Vidal, Phys. Rev. Lett. {\bf 93}, 040502 (2004).
\bibitem{cirac} F. Verstraete, V. Murg, and J. I. Cirac, Adv. Phys {\bf 57}, 143 (2010).
\bibitem{cirac2}F. Verstraete, J. J. Garc\'{\i}a-Ripoll, and J. I. Cirac, Phys. Rev. Lett. {\bf 93}, 207204 (2004); M. Zwolak and G. Vidal, Phys. Rev. Lett. {\bf 93}, 207205
(2004).

\end{thebibliography}
\end{document}